\documentclass[preprint,showpacs,preprintnumbers,amsmath,amssymb]{revtex4}
\usepackage{graphics,epsfig,subfigure}
\usepackage{color}

\begin{document}
\newcommand{\PR}[1]{\ensuremath{\left[#1\right]}} 
\newcommand{\PC}[1]{\ensuremath{\left(#1\right)}} 
\newcommand{\PX}[1]{\ensuremath{\left\lbrace#1\right\rbrace}} 
\newcommand{\BR}[1]{\ensuremath{\left\langle#1\right\vert}} 
\newcommand{\KT}[1]{\ensuremath{\left\vert#1\right\rangle}} 
\newcommand{\MD}[1]{\ensuremath{\left\vert#1\right\vert}} 

\title{Thick brane in reduced Horndeski theory}
\author{Qi-Ming Fu$^{1,}$\footnote{fuqiming@snut.edu.cn},
        Hao Yu$^{2,}$\footnote{yuh13@lzu.edu.cn},
        Li Zhao$^{2,}$\footnote{lizhao@lzu.edu.cn},
        and Yu-Xiao Liu$^{2,}$\footnote{liuyx@lzu.edu.cn, corresponding author}}

\affiliation{$^{1}$Institute of Physics, Shannxi University of Technology, Hanzhong 723000, China \\
             $^{2}$Institute of Theoretical Physics $\&$ Research Center of Gravitation, Lanzhou University, Lanzhou 730000, China}

\begin{abstract}
Horndeski theory is the most general scalar-tensor theory retaining second-order field equations, although the action includes higher-order terms. This is achieved by a special choice of coupling constants. In this paper, we investigate thick brane system in reduced Horndeski theory, especially the effect of the non-minimal derivative coupling on thick brane. First, the equations of motion are presented and a set of analytic background solutions are obtained. Then, to investigate the stability of the background scalar profile, we present a novel canonically normalized method, and show that although the original background scalar field is unstable, the canonical one is stable. The stability of the thick brane under tensor perturbation is also considered. It is shown that the tachyon is absent and the graviton zero mode can be localized on the brane. The localized graviton zero mode recovers the four-dimensional Newtonian potential and the presence of the non-minimal derivative coupling results in a splitting of its wave function. The correction of the massive graviton KK modes to the Newtonian potential is also analyzed briefly.
\end{abstract}





\maketitle

\section{Introduction}

It is well known that general relativity is just an effective gravitational theory at low energy and at the scale of the Solar System because of its nonrenormalization and incapable of explaining dark matter and dark energy. Thus, it needs to be modified at high energy and galactic scale, such as adding higher-order curvature terms and introducing extra scalar fields. In this paper, we mainly focus on the most general scalar-tensor gravitational theory, i.e., Horndeski theory \cite{Horndeski:1974wa}, which maintains the second-order equations of motion (for a recent review, see Refs.~\cite{Gao2011,Kobayashi2019} and references therein). The action is
\begin{equation}
S=\int d^4x\sqrt{-g}\left(\sum_{i=2}^5L_i\right),~\label{horndeskieq1}
\end{equation}
where
\begin{eqnarray}
L_2&=&K(\phi,X), \\
L_3&=&-G_3(\phi,X)\Box \phi,\\
L_4&=&G_4(\phi,X)R+G_{4,X}(\phi,X)\left[(\Box\phi)^2-(\nabla_\mu\nabla_\nu\phi)(\nabla^\mu\nabla^\nu\phi)\right],\\
L_5&=&G_5(\phi,X)G_{\mu\nu}\nabla^\mu\nabla^\nu\phi-\frac{1}{6}G_{5,X}(\phi,X)\big[(\Box\phi)^3-3(\Box\phi)(\nabla_\mu\nabla_\nu\phi)(\nabla^\mu\nabla^\nu\phi) \nonumber\\
   &+&2(\nabla^\mu\nabla_\alpha\phi)(\nabla^\alpha\nabla_\beta\phi)(\nabla^\beta\nabla_\mu\phi)\big].
\end{eqnarray}
Here $\Box\phi=\nabla_\alpha\nabla^\alpha\phi$, $G_{\mu\nu}= R_{\mu\nu}-\frac{1}{2}g_{\mu\nu}R$, $K$, $G_i$ ($i=3,4,5$) are functions of the scalar field $\phi$ and its kinetic term $X=-\nabla_\mu\phi\nabla^\mu\phi/2$, and $G_{i,X}(\phi,X)=\partial G_i(\phi,X)/\partial X$.

Provided $K(\phi,X)=X-V(\phi)$, $G_4(\phi,X)=M_{\text{Pl}}^2/2$, and $G_3$, $G_5$ are just functions of $\phi$ but no more than quadratic, the action (\ref{horndeskieq1}) can be largely simplified as
\begin{eqnarray}
S=\int d^4x \sqrt{-g}\Big[\frac{M_{\text{Pl}}^2}{2}R+(b_0+b_1\phi)G_{\mu\nu}\nabla^{\mu}\nabla^{\nu}\phi-(c_0+c_1\phi)\Box\phi+X-V(\phi)\Big],~\label{L1}
\end{eqnarray}
where $M_{\text{Pl}}$ is the Planck scale and $b_0,~b_1,~c_0$, $c_1$ are some constants.

After integrating by parts and neglecting surface terms $b_0\nabla^{\mu}(G_{\mu\nu}\nabla^{\nu}\phi)$ and $c_0\Box \phi$, the action (\ref{L1}) becomes
\begin{eqnarray}
S=\int d^4x \sqrt{-g}\left[\frac{M_{\text{Pl}}^2}{2 }R-b_1 G_{\mu\nu}\nabla^{\mu}\phi\nabla^{\nu}\phi-\left(\frac{1}{2}-c_1\right)g^{\mu\nu}\nabla_\mu\phi\nabla_\nu\phi-V(\phi) \right].
\end{eqnarray}
Defining $\tilde{\phi}\equiv\sqrt{1-2c_1}\phi$ with $c_1\leqslant 1/2$, one obtains
\begin{eqnarray}
S=\int d^4x \sqrt{-g}\left[\frac{M_{\text{Pl}}^2}{2 }R-b G_{\mu\nu}\nabla^\mu\tilde{\phi}\nabla^\nu\tilde{\phi}-\frac{1}{2}g^{\mu\nu}\nabla_\mu\tilde{\phi}\nabla_\nu\tilde{\phi}-V(\tilde{\phi}) \right],~\label{NMD}
\end{eqnarray}
where $b\equiv\frac{b_1}{1-2c_1}$, and the second term is the so-called non-minimal derivative coupling \cite{Germani2010}.
Actually, the combination of $\kappa_1 R g_{\mu\nu} \nabla^{\mu}\phi\nabla^{\nu}\phi$ and $\kappa_2 R_{\mu\nu}\nabla^{\mu}\phi\nabla^{\nu}\phi$ is the most general term that the gravitational sector couples to the kinetic term of the scalar field in scalar-curvature coupling theory \cite{Capozziello2000}, and it was shown in Ref.~\cite{Sushkov2009} that the equations of motion of the field $g_{\mu\nu}$ and $\phi$ reduce to second order if $\kappa=\kappa_2=-2\kappa_1$, which leads to the non-minimal derivative coupling between the Einstein tensor and scalar field,  $\kappa G_{\mu\nu}\nabla^{\mu}\phi\nabla^{\nu}\phi$.

Theories of the type (\ref{NMD}) have been extensively investigated in different contexts. For instance, the conditions of existing a stable Einstein static universe under both scalar and tensor perturbations were investigated in Refs.~\cite{Atazadeh2015,Huang2018}. The quasi-normal modes of asymptotic anti-de Sitter (AdS) black holes were explored in Ref.~\cite{Dong2017}. In Ref.~\cite{Miao2016}, the authors investigated thermodynamic properties of a new class of black holes in these theories and showed that this class of black holes presents rich thermodynamic behaviors and critical phenomena. In Ref.~\cite{Feng2019}, the authors calculated the holographic complexity of AdS black holes in these theories and showed that the action growth for planar and spherical topologies satisfies the Lloyd's bound. Besides, the non-minimal derivative coupling can result in the present cosmic acceleration, and provide an inflationary mechanism \cite{Capozziello2000,Amendola1993,Capozziello1999,Tsujikawa2012,Sadjadi2014,Darabi2015,Yang2016}. For a recent review, see Ref.~\cite{Gumjudpai2015}.

On the other hand, inspired by string theory, braneworld models draw much attention in recent years. There are mainly two kinds of braneworld models: thin braneworld model and thick braneworld model. The most famous thin braneworld models are the Randall-Sundrum (RS) type models \cite{rs1,rs2} and their extensions \cite{Davoudiasl2000,Davoudiasl2001,Gherghetta2001,Huber2001}. These models were extensively investigated in the last decades because of their advantages in solving some long-existing problems, such as the gauge hierarchy problem, the fermion mass hierarchy, the cosmological constant problem, and so on. The thick braneworld model was first introduced in Ref.~\cite{DeWolfe2000} and further developed in Refs.~\cite{Csaki2000,Gremm2000a}. In thick brane models, the size of the extra dimension is usually infinity, and the brane is generated dynamically instead of introduced by hand and the energy density of the brane is replaced with a smooth function along the extra dimension \cite{DeWolfe2000,Csaki2000,Gremm2000a,Dzhunushaliev2008,Dzhunushaliev2009,Dzhunushaliev2010a,Liu2011,Giovannini2001,Charmousis2003,Bazeia2004a,Minamitsuji2006,Bazeia2015} instead of a delta function in the thin braneworld models. For a brief review, see Ref.~\cite{Liu2017}. Except the above advantages, the thick branewolrd models also have the following promising features: (1) the effective low energy theory includes a localized massless graviton which can be used to produce the four-dimensional Newtonian potential \cite{Csaki2000, Gremm2000a,Kobayashi2002,Andrianov2008,Barbosa-Cendejas2008,Bazeia2009,Zhong2011,Andrianov2013,Barbosa-Cendejas2005,Veras2016,Barbosa-Cendejas2014,Cui2018,Farakos2007,Giovannini2001a}; (2) the massless scalar graviton decouples from the brane system which avoids the fifth force \cite{Chen2018,Giovannini2001a,Giovannini011,Giovannini2003}; (3) the localization of the fermion field and its chirality can be guaranteed by introducing an interaction with the background scalar field \cite{Melfo2006,Liu2008,Liu2009,Dantas2015}; (4) the gauge field also can be localized on the brane if a mass term is added \cite{Arai2013,Zhao2015,Vaquera-Araujo2015}; (5) the charge universality of the gauge boson can also be obtained via the Dvali-Shifman mechanism \cite{Dvali1997,Maru2001,Kakizaki2002,Davies2008,Callen2014}. Thus, the thick braneworld model is a kind of candidate model producing both Standard Model and Newtonian gravity, which are two key elements forming our four-dimensional world.

Although the non-minimal derivative coupling has been considered in different context, such as cosmology and black hole, the effect of this term on thick brane is still unclear in the literature. Recently, the thick brane under non-minimal derivative coupled gravity has been studied in Ref.~\cite{Brito2018}. However, the background solutions of the thick brane were only solved numerically and approximately and the non-minimal coupling was not included. Besides, because of the non-minimal derivative coupling, the scalar field in the action is not canonical. To study the stability of the thick brane against the quantum tunneling of the background scalar field, we need to canonically normalize the scalar field to get the canonical one and analyse the effective potential of it. The equation of motion of the tensor perturbation $\bar{h}_{\mu\nu}$ given in Ref.~\cite{Brito2018} is not correct because the factor of $\bar{h}_{\mu\nu}$ should be eliminated for the flat brane after inserting the background equations of motion. These are motivations of our present work.

In Sec.~\ref{fieldeq}, we present the action of the thick brane system and derive the equations of motion. In Sec.~\ref{solution}, a set of analytic solutions are obtained and the stability and canonical normalization of the scalar field are considered. In Sec.~\ref{tensorper}, the stability of the brane system under tensor perturbation, the localization of the graviton zero mode, and a brief analysis about the correction of the graviton KK modes to the Newtonian potential are investigated. Section \ref{conclusion} comes with the conclusion.

\section{Field Equations}~\label{fieldeq}

In this paper, we study the thick brane system with the following action:
\begin{eqnarray}
S=\int d^5x \sqrt{-g}\Big[\frac{1}{2 }F(\phi)R-bG_{MN}\nabla^{M}\phi\nabla^{N}\phi+X-V(\phi)\Big], ~\label{actionf}
\end{eqnarray}
which is a generalization of the action in Ref.~\cite{Germani2010}. Here we have set the 5D Planck mass $M_5=1$.
To avoid ghost propagations, the parameter $b$ should be negative. Varying the above action with respect to the metric $g_{MN}$, we obtain the equations of motion
\begin{eqnarray}
F(\phi)G_{MN}-(\nabla_M\nabla_N F(\phi)-g_{MN}\Box^{(5)} F(\phi))= T_{MN}+2b\Theta_{MN},~\label{EoMofEinstein}
\end{eqnarray}
where $\Box^{(5)}\equiv g^{MN}\nabla_{M}\nabla_{N}$,
\begin{eqnarray}
T_{MN}=\nabla_M\phi\nabla_N\phi-\frac{1}{2}g_{MN}(\nabla\phi)^2-g_{MN}V(\phi),
\end{eqnarray}
and
\begin{eqnarray}
\label{theta}
\begin{split}
\Theta_{MN}=&-\frac12\nabla_M\phi\nabla_N\phi R+2\nabla_K\phi\nabla_{(M}\phi R_{N)}^K\\
&-\frac{1}{2}\left(\nabla\phi\right)^2G_{MN}+\nabla^K\phi\nabla^L\phi R_{MKNL}\\
&+\nabla_M\nabla^K\phi\nabla_N\nabla_K\phi-\nabla_M\nabla_N\phi\Box^{(5)}\phi\\
&+g_{MN}\left[-\frac{1}{2}\nabla^K\nabla^L\phi\nabla_K\nabla_L\phi+\frac{1}{2}\left(\Box^{(5)}\phi\right)^2
-\nabla_K\phi\nabla_L\phi R^{KL}\right].
\end{split}
\end{eqnarray}

The equation of motion of the scalar field can be obtained by varying the action (\ref{actionf}) with respect to $\phi$,
\begin{eqnarray}
F_{\phi}R-2  V_{\phi}+2 (g^{KN}+2bG^{KN})\nabla_K\nabla_N\phi=0, \label{scalareom}
\end{eqnarray}
where $F_{\phi}\equiv dF(\phi)/d\phi$ and $V_{\phi}\equiv dV(\phi)/d\phi$.

\section{Brane World Model}~\label{solution}

In general, the line element of a static flat braneworld can be assumed as
\begin{eqnarray}
d^2s=e^{2A(y)}\eta_{\mu\nu}dx^{\mu}dx^{\nu}+dy^2,~\label{le}
\end{eqnarray}
with $y$ the extra coordinate. The warp factor $A$ and scalar field $\phi$ are independent of the brane coordinates, i.e., $A=A(y)$ and $\phi=\phi(y)$.
Then, Eqs.~(\ref{EoMofEinstein}) and (\ref{scalareom}) with the ansatz (\ref{le}) can be reduced to
\begin{eqnarray}
-6 b    A'' \phi '^2+6 A' \left(F'-2 b    \phi ' \phi ''\right)-12 b    A'^2 \phi '^2 \nonumber\\
+6 F \left(A''+2 A'^2\right)+2 F''+2    \text{V}+   \phi '^2&=&0, \label{eom1} \\
-36 b    A'^2 \phi '^2+8 A' F'+12 F A'^2+2   \text{V}-   \phi '^2&=&0, \label{eom2} \\
(1+12bA'^2)\phi''+\left[1+b(6A''+12A'^2)\right]4A'\phi' \nonumber\\
-\left[V_{\phi}+(4A''+10A'^2)F_{\phi}\right]&=&0, \label{scalareom1}
\end{eqnarray}
where the prime denotes the derivative with respect to the extra dimension $y$.
The dynamics of the brane system is determined by Eqs.~(\ref{eom1}), (\ref{eom2}) and (\ref{scalareom1}), but only two of them are independent. However, there are four independent variables, i.e., $A(y)$, $\phi(y)$, $F(\phi(y))$ and $V(\phi(y))$. Thus, to solve this system, we consider the following warp factor and scalar field
\begin{eqnarray}
A(y)&=&\text{ln}[\text{sech}(k y)], \label{warpfactor} \\
\phi(y)&=&v \text{tanh}(k y). \label{scalar}
\end{eqnarray}

\begin{figure*}[htb]
\begin{center}
\subfigure[$F(\phi)$]  {\label{Fyf}
\includegraphics[width=4cm]{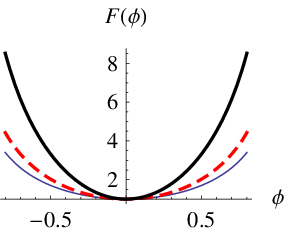}}
\subfigure[$V(\phi)$]  {\label{Vphi}
\includegraphics[width=4cm]{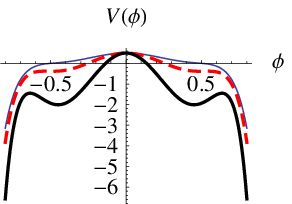}}
\end{center}
\caption{Plots of the brane solutions $F(\phi)$ and $V(\phi)$. The parameters are set to $k=v=1$, $b=-1$ for blue lines, $b=-1.8$ for red dashed thick lines, $b=-5$ for black thick lines.}
\label{A}
\end{figure*}

Now, the system can be solved as
\begin{eqnarray}
F(\phi(y))&=&\frac{1}{399} \Big[   v^2 \left(63 b k^2 \text{sech}^4(k y)+3 \left(36 b k^2+19\right) \text{sech}^2(k y)+72 b k^2+38\right) \nonumber\\
&&+\left(399-   v^2 \left(243 b k^2+95\right)\right) \cosh \left(\theta(y)\right)\Big], ~\label{Fy} \\
V(\phi(y))&=&\frac{k^2}{3192   } \Big[-3    v^2 \text{sech}^6(k y) \big(-3 \left(2588 b k^2+95\right) \cosh (2 k y) \nonumber\\
&&+16 \left(36 b k^2+19\right) \cosh (4 k y)+\left(36 b k^2+19\right) \cosh (6 k y)+7152 b k^2-570\big) \nonumber\\
&&+48 \left(   v^2 \left(243 b k^2+95\right)-399\right) \tanh ^2(k y)\cosh \left(\theta(y)\right) \nonumber\\
&&-32 \sqrt{3} \left(   v^2 \left(243 b k^2+95\right)-399\right) \tanh (k y) \text{sech}(k y) \sinh \left(\theta(y)\right)\Big],~\label{Vy}
\end{eqnarray}
where $\theta(y)\equiv \left(2 \sqrt{3} \tan ^{-1}\left(\tanh \left(ky/2\right)\right)\right)$.
The above two functions can also be expressed in terms of the scalar field $\phi$ as
\begin{eqnarray}
F(\phi)&=&\frac{1}{399 v^2}\bigg[63 b \phi^4-3 (78 b+19) \phi^2 v^2+(243 b+95) v^4 \nonumber\\
&& +v^2 \left(399-(243 b+95) v^2\right)\cosh \left(\theta(\phi)\right)\bigg],~\label{Fphi0} \\
V(\phi)&=&\frac{1}{798 v^4}\bigg[11592 b \phi ^6+105 (19-192 b) \phi ^4 v^2+6 (1284 b-475) \phi ^2 v^4+399 v^6 \nonumber\\
&&+4 \phi v^2 \left((243 b+95) v^2-399\right) \bigg(3 \phi  \cosh \left(\theta(\phi)\right) -2 v \sqrt{3-\frac{3 \phi ^2}{v^2}} \sinh \left(\theta(\phi)\right)\bigg)\bigg],~\label{Vphi0}
\end{eqnarray}
where $\theta(\phi)\equiv 2 \sqrt{3} \tan ^{-1}\left(\tanh \left((1/2) \tanh ^{-1}\left(\phi/v\right)\right)\right)$ and we have chosen $k=1$. Figure \ref{Fyf} shows that the value of $F(\phi)$ is always positive which is a requirement of the positive definiteness of the coefficient of the scalar curvature. Figure \ref{Vphi} shows that two metastable states appear with the decreasing of the parameter $b$. However, the potential $V(\phi)$ is bottomless which seemingly indicates an instability of the scalar profile. To be specific, although the configuration of the background scalar field is a kink, this profile will be changed drastically if considering the quantum tunneling. Then, the configuration of the thick brane will be changed.
However, the situation is ambiguous because of two reasons: (1) the kinetic term of the scalar field is not canonical; (2) the potential of the scalar field is determined not only by $V(\phi)$ but also by the non-minimal coupling $F(\phi)R$. Thus, it is necessary to reconsider this problem carefully.

Equation (\ref{scalareom1}) can be rewritten as
\begin{eqnarray}
\phi''+4\tilde{A}'\phi'-\frac{\partial V_{\text{eff}}(\phi)}{\partial\phi}=0, \label{scalareom2}
\end{eqnarray}
where
\begin{eqnarray}
\tilde{A}&\equiv& \int A'\left(1+\frac{6bA''}{12bA'^2+1}\right)dy, \\
V_{\text{eff}}&\equiv& \int \frac{4F'A''+10A'^2F'+  V'}{ (12bA'^2+1)}dy,
\end{eqnarray}
which indicates the kinetic term of the scalar field $\phi$ is canonical in an effective spacetime with the warp factor $\text{e}^{2\tilde{A}}$.
To obtain a canonical kinetic term in the physical spacetime, we define a new scalar field
\begin{eqnarray}
\tilde{\phi}\equiv \int \text{e}^{4(\tilde{A}-A)}\phi' dy.
\end{eqnarray}
Substituting $\tilde{\phi}$ into Eq.~(\ref{scalareom2}), one can obtain
\begin{eqnarray}
\tilde{\phi}''+4A'\tilde{\phi}'-\frac{\partial\tilde{V}_{\text{eff}}(\tilde{\phi})}{\partial\tilde{\phi}}=0,
\end{eqnarray}
where
\begin{eqnarray}
\tilde{V}_{\text{eff}}\equiv \int \frac{\tilde{g}}{g}\partial_yV_{\text{eff}}~dy,
\end{eqnarray}
and $\tilde{g}$ is the determinant of the effective spacetime metric $ds^2=\text{e}^{2\tilde{A}}\eta_{\mu\nu}dx^{\mu}dx^{\nu}+dy^2$.
Thus, the action of the scalar field after being canonically normalized can be written as
\begin{eqnarray}
S=\int dx^5 \sqrt{-g}\left(-\frac{1}{2}g^{MN}\partial_{M}\tilde{\phi}\partial_{N}\tilde{\phi}-\tilde{V}_{\text{eff}}(\tilde{\phi})\right),
\end{eqnarray}
which indicates that the canonical background scalar field generating the thick brane should be $\tilde{\phi}$ but not $\phi$, and the stability of the configuration of the thick brane should be determined by the effective potential $\tilde{V}_{\text{eff}}(\tilde{\phi})$. By the way, the above analysis is also an effective way to canonically normalize a field with non-standard kinetic term. With Eqs.~(\ref{warpfactor}) and (\ref{scalar}), the warp factor of the effective spacetime and the canonical background scalar field can be calculated as
\begin{eqnarray}
\tilde{A}(y)&=&\frac{1}{4} \ln \Big(\text{sech}^4(k y) \left(12 b k^2 \text{tanh}^2(k y)+1\right)\Big), \\
\tilde{\phi}(y)&=&v \tanh (k y) \left(4 b k^2 \tanh ^2(k y)+1\right).
\end{eqnarray}

\begin{figure*}[htb]
\begin{center}
\subfigure[$b=-0.5$]  {\label{Vefft1}
\includegraphics[width=4cm]{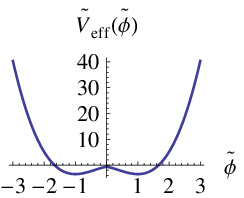}}
\subfigure[$b=-1$]  {\label{Vefft2}
\includegraphics[width=4cm]{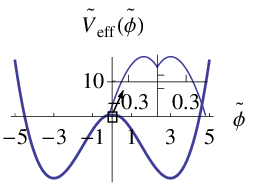}}
\end{center}
\caption{Plots of the effective potential $\tilde{V}_{\text{eff}}(\tilde{\phi})$. The parameters are set to $k=v=1$.}
\label{B}
\end{figure*}

It can be easily shown that $\tilde{A}=A$ and $\tilde{\phi}=\phi$ when $b=0$, which justifies the above analysis. It is obvious that $\tilde{\phi}(\pm\infty)\rightarrow \pm v(4b k^2+ 1)$, which indicates that there is a critical value $b_c =-\frac{1}{4k^2}$.
When $b_c<b\leqslant 0$, the scalar field is a kink, and when $b<b_c$, the scalar field is an anti-kink. Figure \ref{A} shows that the scalar field $\tilde{\phi}(y)$ changes from a kink to an anti-kink with $|b|$ increasing,
and Fig. \ref{B} plots its effective potential. It is obvious that there are always two stable minimums in the effective potential $\tilde{V}_{\text{eff}}(\tilde{\phi})$. Thus, the profile of the canonical background scalar field is stable, so as the configuration of the thick brane. Besides, with $|b|$ increasing, another metastable state around the origin appears, which may indicate an inner structure of the thick brane.

\begin{figure*}[htb]
\begin{center}
\subfigure[$\tilde{\phi}(y)$]  {\label{phity1}
\includegraphics[width=4cm]{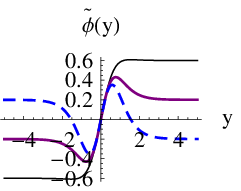}}
\subfigure[$\tilde{\phi}(y)$]  {\label{phity2}
\includegraphics[width=4cm]{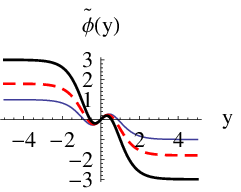}} \\
\end{center}
\caption{Plots of the canonical background scalar field $\tilde{\phi}(y)$. The parameters are set to $k=v=1$, $b=-0.1$ for black thin line, $b=-0.2$ for purple thick line, $b=-0.3$ for blue dashed thick line, $b=-0.5$ for blue thin line, $b=-0.7$ for red dashed thick line, $b=-1$ for black thick line.}
\label{A}
\end{figure*}

\section{Tensor Perturbation}~\label{tensorper}

In general, the tensor, vector and scalar perturbations are decoupled from each other. Thus, they can be investigated individually. The metric under the tensor perturbation can be written as
\begin{eqnarray}
ds^2=\text{e}^{2A(y)}(\eta_{\mu\nu}+h_{\mu\nu})dx^{\mu}dx^{\nu}+dy^2, \label{flumetric}
\end{eqnarray}
where $h_{\mu\nu}$ represents the transverse and traceless (TT) tensor perturbation, i.e., $\eta^{\mu\alpha}\partial_{\alpha}h_{\mu\nu}=0$ and $h\equiv \eta^{\mu\nu}h_{\mu\nu}=0$. Then, the perturbation equation can be calculated as
\begin{eqnarray}
&F&\delta G_{\mu\nu}-(A'\text{e}^{2A}h_{\mu\nu}+\frac{1}{2}\text{e}^{2A}h'_{\mu\nu})F'+\text{e}^{2A}h_{\mu\nu}\Box^{(5)}F
-\eta_{\mu\nu}h^{\rho\sigma}\nabla_{\rho}\nabla_{\sigma}F \nonumber\\
&=&- \text{e}^{2A}h_{\mu\nu}\left(\frac{1}{2}\phi'^2+V\right)
+2b \delta\Theta_{\mu\nu},
\end{eqnarray}
where
\begin{eqnarray}
\delta\Theta_{\mu\nu}&=&-\frac{1}{2}\phi'^2\delta G_{\mu\nu}-\text{e}^{-2A}h^{\rho\sigma}\nabla_{\nu}\nabla_{\rho}\phi\nabla_{\mu}\nabla_{\sigma}\phi+2(\nabla_{\mu}\nabla^{\sigma}\phi)(\phi'A'\text{e}^{2A}h_{\nu\sigma}+\frac{1}{2}\phi'\text{e}^{2A}h'_{\nu\sigma}) \nonumber\\
&+&\text{e}^{-2A}h^{\rho\sigma}\nabla_{\rho}\nabla_{\sigma}\phi\nabla_{\mu}\nabla_{\nu}\phi-\Box^{(5)}\phi(\phi'A'\text{e}^{2A}h_{\mu\nu}+\frac{1}{2}\phi'\text{e}^{2A}h'_{\mu\nu}) \nonumber\\
&+&\phi'^2(-A''\text{e}^{2A}h_{\mu\nu}-A'^2\text{e}^{2A}h_{\mu\nu}-A'\text{e}^{2A}h'_{\mu\nu}-\frac{1}{2}\text{e}^{2A}h''_{\mu\nu}) \nonumber\\
&+&\text{e}^{2A}h_{\mu\nu}\left(\frac{1}{2}\nabla^{T}\nabla^{L}\phi\nabla_{T}\nabla_{L}\phi+\frac{1}{2}\Box^{(5)}\phi\Box^{(5)}\phi-R_{KL}\nabla^{K}\phi\nabla^{L}\phi\right) \nonumber\\
&+&\text{e}^{2A}\eta_{\mu\nu}\Big(\text{e}^{-2A}h^{\rho\sigma}(\nabla_{\sigma}\nabla^{\tau}\phi+\nabla_{\sigma}\phi'\nabla_{\rho}\phi') \nonumber\\
&-&(\nabla^{\rho}\nabla^{\sigma}\phi)(\phi'A'\text{e}^{2A}h_{\rho\sigma}+\frac{1}{2}\phi'\text{e}^{2A}h'_{\rho\sigma})-\text{e}^{-2A}h^{\rho\sigma}\nabla_{\rho}\nabla_{\sigma}\phi\Box^{(5)}\phi\Big).
\end{eqnarray}

Considering the TT conditions, the above equation can be reduced to
\begin{eqnarray}
T(y)h''_{\mu\nu}+B(y)h'_{\mu\nu}+\text{e}^{-2A}\Box^{(4)}h_{\mu\nu}=0, \label{fluy}
\end{eqnarray}
where $\Box^{(4)}\equiv \eta^{\mu\nu}\partial_{\mu}\partial_{\nu}$, and
\begin{eqnarray}
T(y)&=&\frac{F-b \phi'^2}{F+b \phi'^2}, \\
B(y)&=&\frac{4A'F+F'-4b \phi'^2A'-2b \phi'\phi''}{F+b \phi'^2}.
\end{eqnarray}
After a coordinate transformation $dy=\text{e}^{A}dz$, Eq.~(\ref{fluy}) becomes
\begin{eqnarray}
N(z)\partial_z^2h_{\mu\nu}+L(z)\partial_z h_{\mu\nu}+\Box^{(4)}h_{\mu\nu}=0, \label{fluz}
\end{eqnarray}
where
\begin{eqnarray}
N(z)&=&\frac{F-b \text{e}^{-2A}(\partial_z\phi)^2}{F+b \text{e}^{-2A}(\partial_z\phi)^2}, ~\label{N}\\
L(z)&=&\frac{3 F \partial_z A+ \partial_z F-b \text{e}^{-2A}(\partial_z A)(\partial_z\phi)^2-2b \text{e}^{-2A}(\partial_z\phi)(\partial_z^2\phi)}{F+b \text{e}^{-2A}(\partial_z\phi)^2}.
\end{eqnarray}

With a further coordinate transformation $dz=\sqrt{N}dw$, Eq.~(\ref{fluz}) can be transformed as
\begin{eqnarray}
\partial_w^2h_{\mu\nu}+\left(\frac{L}{\sqrt{N}}-\frac{\partial_w N}{2N}\right)\partial_w h_{\mu\nu}+\Box^{(4)}h_{\mu\nu}=0.~\label{fluw}
\end{eqnarray}
Considering the decomposition $h_{\mu\nu}(x,w)=\varepsilon_{\mu\nu}(x)\text{e}^{-ipx}H(w)$ with $p^2=-m^2$, this equation simplifies to
\begin{eqnarray}
\partial_w^2H+Q(w)\partial_w H+m^2 H=0,
\end{eqnarray}
where $Q(w)=\left(\frac{L}{\sqrt{N}}-\frac{\partial_w N}{2N}\right)$.
Then, by redefining $H(w)=G(w)\tilde{H}(w)$ with $G(w)= \text{exp}(-\frac{1}{2}\int Q(w)dw)$, we can obtain a Schr$\ddot{\text{o}}$dinger-like equation:
\begin{eqnarray}
-\partial_w^2 \tilde{H} + U(w)\tilde{H} = m^2 \tilde{H},
\end{eqnarray}
where $U(w)=\left(\frac{1}{2}\partial_w Q+\frac{1}{4}Q^2\right)$.
The above equation can be factorized as
\begin{eqnarray}
\left(\partial_w+\frac{Q}{2}\right)\left(-\partial_w+\frac{Q}{2}\right)\tilde{H}=m^2 \tilde{H},~\label{flusy}
\end{eqnarray}
which indicates there is no tachyon state, i.e., $m^2\geq 0$. Thus, the brane is stable under the tensor perturbation.

By setting $m=0$, the graviton zero mode can be solved from Eq.~(\ref{flusy}):
\begin{eqnarray}
\tilde{H}_0=N_0\text{exp}\left(\frac{1}{2}\int Q dw\right)=N_0\text{exp}\left(\frac{1}{2}\int \frac{Q}{\sqrt{N}} dz\right),~\label{zeromode}
\end{eqnarray}
where $N_0$ is a normalization constant. The normalization condition of the graviton zero mode is
\begin{eqnarray}
\int \tilde{H}_0^2(w) dw=\int \tilde{H}_0^2(z)\frac{dz}{\sqrt{N(z)}}<\infty.
\end{eqnarray}

\subsection{Localization of the graviton zero mode}
\begin{figure*}[htb]
\begin{center}
\subfigure[$U(w)$]  {\label{Uw}
\includegraphics[width=4cm]{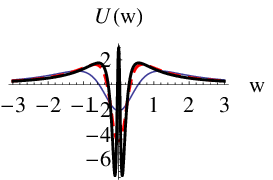}}
\subfigure[$H_0(w)$]  {\label{H0w}
\includegraphics[width=4cm]{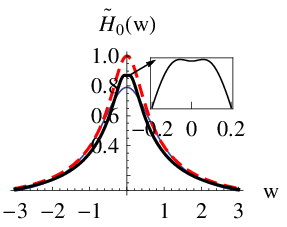}}
\end{center}
\caption{Plots of the effective potential of the graviton and the wave function of the graviton zero mode. The parameters are set to $k=1$, $b=0$ for blue lines, $b=-0.12$ for red dashed thick lines, and $b=-0.125$ for black thick lines.}
\label{g}
\end{figure*}

In the following, to obtain some analytic results we assume the relation $v^2 \left(243 b k^2+95\right)-399=0$. Then,
the graviton zero mode can be solved as
\begin{eqnarray}
\tilde{H}_0(z)&=&N_0\text{exp}\left(\frac{1}{2}\int \frac{Q}{\sqrt{N}} dz\right) \nonumber\\
              &=&N_0\left(k^2 z^2+1\right)^{-7/4}\big[ 2 k^4 z^4 \left(36 b k^2+19\right)+7k^2 z^2 \left(36 b k^2+19\right)-156k^2 b+95\big]^{1/4} \nonumber\\ &&\times\big[2 k^4 z^4 \left(36 b k^2+19\right)+7k^2 z^2 \left(36 b k^2+19\right)+642k^2 b+95\big]^{1/4},
\end{eqnarray}
and the normalization condition can be calculated as
\begin{eqnarray}
\int \tilde{H}_0^2(z)\frac{dz}{\sqrt{N(z)}}&=&N_0^2\int_{-\infty }^{\infty } \frac{6 b k^2 \left(12 k^4 z^4+42 k^2 z^2+107\right)+19 \left(2 k^4 z^4+7 k^2 z^2+5\right)}{\left(k^2z^2+1\right)^{7/2}} dz \nonumber\\
                                           &=&\frac{8 N_0^2 \left(488 b k^2+95\right)}{5 k}=1.~\label{n0}
\end{eqnarray}
Then, $N_0$ can be solved as $N_0=\sqrt{\frac{5k}{8(488bk^2+95)}}$. To ensure $N_0$ is real, the parameter $b$ should satisfy $b>-\frac{95}{488 k^2}$ for $k>0$.
Besides, to ensure the coordinate transformation $dz=\sqrt{N}dw$ is well defined, $N(z)$ should be positive, namely,
\begin{eqnarray}
\frac{\left(36 b k^2+19\right) \left(2 k^4 z^4+7 k^2 z^2\right)-156 b k^2+95}{\left(36 b k^2+19\right) \left(2 k^4 z^4+7 k^2 z^2\right)+642 b k^2+95}>0,
\end{eqnarray}
which gives $-\frac{95}{642 k^2}<b<\frac{95}{156 k^2}$ for $z\in (-\infty,+\infty)$. Thus, to avoid the ghost gravitons and produce the Newtonian potential, the parameter $b$ should satisfy $-\frac{95}{642 k^2}<b\leqslant0$. Figure \ref{g} shows the effective potential of the graviton along the extra dimension and the wave function of the graviton zero mode. When $b=0$, it reduces to a volcano-like potential which is the case of the thick brane in general relativity \cite{Csaki2000}.

Because the integral $w=\int\frac{1}{\sqrt{N(z)}}dz$ is difficult, we can not obtain an analytic relation for $w(z)$. Thus, the graviton zero mode and its effective potential can only be expressed analytically in terms of $z$. However, some behaviors of $\tilde{H}_0(w)$ and $U(w)$ at $w=0$ can be obtained with the relations $\partial_w^2\tilde{H}_0(w)=\frac{1}{2}\partial_z N(z)\partial_z\tilde{H}_0(z)+N(z)\partial_z^2\tilde{H}_0(z)$, $\partial_wU(w)=\sqrt{N(z)}\partial_zU(z)$, and $\partial_w^2U(w)=\frac{1}{2}\partial_z N(z)\partial_zU(z)+N(z)\partial_z^2U(z)$.
Then, we have
\begin{eqnarray}
\partial_w^2U(0)&=&\frac{9 k^4}{2 \left(642 b k^2+95\right)^4}\big(172617477120 b^4 k^8-87248865792 b^3 k^6-22421374992 b^2 k^4 \nonumber\\
                &&+2811366920 b k^2+413769175\big), ~\label{Uzpp0} \\
U(0)&=&\frac{399 k^2 \left(2064 b^2 k^4-528 b k^2-95\right)}{2 \left(642 b k^2+95\right)^2},~\label{behav0} \\
\partial_w^2\tilde{H}_0(0)&=&\frac{399 k^2 \sqrt[4]{95-156 b k^2} \left(2064 b^2 k^4-528 b k^2-95\right)}{2 \left(642 b k^2+95\right)^{7/4}}.~\label{H0z0}
\end{eqnarray}
From Eq.~(\ref{Uzpp0}), $\partial_w^2U(0)<0$ when $-\frac{95}{642 k^2}<b<-\frac{0.1032}{k^2}$, which indicates a double-well potential appears. From Eqs.~(\ref{behav0}) and (\ref{H0z0}), $\partial_w^2\tilde{H}_0(0)<0$ and $U(0)>0$ when $-\frac{95}{642 k^2}<b<-\frac{0.1218}{k^2}$, which indicates the splitting of the graviton zero mode.

In conclusion, the effective potential of the graviton is volcano-like when $b\rightarrow 0$, and changes to double-well when $-\frac{95}{642 k^2}<b<-\frac{0.1032}{k^2}$. Besides, the wave function of the graviton zero mode splits when $-\frac{95}{642 k^2}<b<-\frac{0.1218}{k^2}$.

\subsection{Correction to the Newtonian potential}
Except for the localized graviton zero mode, there are a lot of continuous massive KK gravitons. These KK modes may lead a correction to the Newtonian potential. In the following, we will give a brief analysis about this.

To obtain an approximate analytic relation for $w(z)$, we considering the thin-brane limit, i.e., $k\gg1$. In this limit, the parameter $b$ should be small. Then, the integrand of $w=\int\frac{1}{\sqrt{N(z)}}dz$ can be expanded in terms of $b$ as
\begin{eqnarray}
\frac{1}{\sqrt{N(z)}}=1+\frac{21}{2 k^4 z^4+7 k^2 z^2+5}b k^2-\frac{63 \left(48 k^4 z^4+168 k^2 z^2+29\right)}{38 \left(2 k^4 z^4+7 k^2 z^2+5\right)^2}(bk^2)^2+\mathcal{O}(b^3).
\end{eqnarray}
With $-\frac{95}{642}<b k^2\leqslant0$ in mind, it can be easily shown that the integrand is dominated by the first two terms of the above expression no matter in the small or large $z$ region. Then, the integral can be integrated out as
\begin{eqnarray}
w(z)=z+7 b k \tan^{-1}(k z)-7 \sqrt{\frac{2}{5}} b k \tan^{-1}\left(\sqrt{\frac{2}{5}} k z\right)+\mathcal{O}(b^2).
\end{eqnarray}
It is obvious that $w\simeq z$ in the large $z$ region. Thus, it is well-approximated to investigate the asymptotic behavior of the effective potential of the graviton in $z$ coordinate,
\begin{eqnarray}
U(z)&=&\frac{15 k^2 z^2 \left(4 k^6 z^6+28 k^4 z^4+37 k^2 z^2-14\right)}{4 z^2 \left(k^2 z^2+1\right)^2 \left(2 k^2 z^2+5\right)^2} \nonumber\\
    &&-\frac{63  \left(k^4 z^4 \left(556 k^6 z^6+3752 k^4 z^4+4909 k^2 z^2-1260\right)\right)}{38 z^4 \left(\left(k^2 z^2+1\right)^3 \left(2 k^2 z^2+5\right)^3\right)}b+\mathcal{O}(b^2) \nonumber\\
    &=&\frac{-\frac{210}{(k z)^6}+\frac{555}{(k z)^4}+\frac{420}{(k z)^2}+60}{4 \left(\frac{25}{(k z)^8}+\frac{70}{(k z)^6}+\frac{69}{(k z)^4}+\frac{28}{(k z)^2}+4\right) z^2} \nonumber\\
    &&-\frac{ \left(-\frac{79380}{(k z)^8}+\frac{309267}{(k z)^6}+\frac{236376}{(k z)^4}+\frac{35028}{(k z)^2}\right)}{38 \left(\frac{125}{(k z)^{12}}+\frac{525}{(k z)^{10}}+\frac{885}{(k z)^8}+\frac{763}{(k z)^6}+\frac{354}{(k z)^4}+\frac{84}{(k z)^2}+8\right) z^4}b+\mathcal{O}(b^2).
\end{eqnarray}
It is obvious that $U(z)\sim \frac{15}{4z^2}$ as $|z|\gg 1$ and $k\gg 1$, which has the particular form $\alpha(\alpha+1)/z^2$. Then, the KK modes for small masses on the brane obey the relation $\psi_m(0)\sim m^{\alpha-1}$ shown in Ref.~\cite{Csaki2000}, and the correction to the Newtonian potential between two massive objects at a distance of $r$ is $\Delta V(r)\varpropto 1/r^{2\alpha}$. For our potential, $\alpha=3/2$, this leads to $|\psi_m(0)|^2\sim m$. Thus the correction to the Newtonian potential is $\Delta V(r)\varpropto 1/r^3$.

\section{Conclusions and Discussion}\label{conclusion}

In this paper, we investigated thick brane system in reduced Horndeski theory, especially the effect of the non-minimal derivative coupling on the thick brane model. Except for the non-minimal derivative coupling, the background scalar field $\phi$ is also non-minimally coupled with the curvature. A set of analytic solutions for the brane system were obtained. It seems that this scalar field is unstable, which means an instability for the brane, because its potential is bottomless. However, because of the non-minimal coupling and the non-minimal derivative coupling, the kinetic term of the scalar field $\phi$ is not canonical and the effective potential of it not only comes from the potential in the action but also from the non-minimal coupling. To obtain the canonical background scalar field and its effective potential, we introduced a novel method, which is independent of specific model.
It was found that the effective potential of the canonical background scalar field $\tilde{V}_{eff}(\tilde{\phi})$ is Mexican-hat-like, which has two stable vacuums thus stable. Besides, the canonical background scalar field $\tilde{\phi}$ changes from kink to double-kink then to anti-kink with $|b|$ increasing. It indicates that the non-minimal derivative coupling may change the inner structure of the brane, which will affect the property of the graviton zero mode and other matter fields.

For tensor perturbation, a Schr$\ddot{\text{o}}$dinger-like equation of the graviton was obtained and its Hamiltonian can be factorized, which ensures the stability of the tensor perturbation of the brane system.
The effective potential of the graviton is volcano-like when $-\frac{0.1032}{k^2}<b\leqslant 0$,
and a double-well structure shows up when $-\frac{95}{642 k^2}<b<-\frac{0.1032}{k^2}$. When $-\frac{95}{642 k^2}<b<-\frac{0.1218}{k^2}$, the wave function of the graviton zero mode splits.
Except for the localized graviton zero mode, there are a lot of continuous KK modes which decouple from the brane system.
Even so, an enough number of massive continuous KK modes will produce an observable correction to the Newtonian potential. We gave a brief analysis about it and found their correction is $\Delta V(r)\varpropto 1/r^3$. The effect of the non-minimal derivative coupling on the scalar perturbation and the correction of their KK modes to the Newtonian potential are also interesting problems. These are left for our future works.

\acknowledgments{
This work was supported by the National Natural Science Foundation of China (Grants No. 11875151 and No. 11522541), and Scientific Research Program Funded by Shaanxi Provincial Education Department (No. 19JS010).
}

\end{document}